\documentclass[a4paper,11pt]{extarticle}  

\usepackage{stylesheet}

\usepackage[left=1in,top=1in,bottom=.65in,right=1in]{geometry}

\title{Leveraging Climate Services to Build Climate Resilient Power Systems}

\author{%
\textbf{%
Laurent Dubus \orcidlink{0000-0002-3987-646X}\textcolor{Accent}{\textsuperscript{1,2 *}}, %
Alberto Troccoli \orcidlink{0000-0001-5870-4367}\textcolor{Accent}{\textsuperscript{2,3}}, %
Aron Zuiker\textcolor{Accent}{\textsuperscript{4}}, %
Laurens Stoop \orcidlink{0000-0003-2756-5653}\textcolor{Accent}{\textsuperscript{5}} %
}\\[0.5em]
\begin{small}
\textcolor{Accent}{\textsuperscript{1}}RTE, Paris, France\\[0.5em] 
\textcolor{Accent}{\textsuperscript{2}}World Energy \& Meteorological Council, UK\\[0.5em] 
\textcolor{Accent}{\textsuperscript{3}}Inside Climate Service, Padua, Italy\\[0.5em] 
\textcolor{Accent}{\textsuperscript{4}}ENTSO-E, Brussels, Belgium\\[0.5em] 
\textcolor{Accent}{\textsuperscript{5}}TenneT TSO B.V., Arnhem, the Netherlands\\[0.5em] 
\textcolor{Accent}{\textsuperscript{*}}Corresponding Author: \textcolor{Accent}{laurent.dubus@rte-france.com} \\[0.5em]
\textcolor{Accent}{Paper submitted to CIGRE Paris Session 2026, C3 PS2 10894} \end{small}
}

\date{}

\begin{document}

% Title page
\maketitle

%%%%%%%%%%%%%%%%%%%%%%%%%%%%%%%%%%%%%%%%%%%%%%%%%%
%%%%		~~~~ Abstract ~~~~
%%%%%%%%%%%%%%%%%%%%%%%%%%%%%%%%%%%%%%%%%%%%%%%%%%
\begin{abstract}
We explore the crucial interplay between climate change and power system planning, highlighting the urgent need to systematically integrate climate information into energy system studies. 
With the global energy sector responsible for roughly two-thirds of greenhouse gas emissions, Europe's strategy focuses on rapid decarbonization and high renewable energy penetration to meet the Paris Agreement and ambitious regional climate targets. 
However, replacing fossil fuels with renewables like wind, solar, and hydropower increases power systems' reliance on weather and climate, creating new vulnerabilities for grid stability and resilience.

Climate change impacts the energy sector on multiple fronts. 
Short-term weather variability drives daily and seasonal fluctuations in supply and demand. 
Long-term trends and increased frequency of extremes pose risks to infrastructure performance, asset lifetimes, and system adequacy. 
Representing compound events and spatial correlations across borders is a complex challenge, and uncertainties persist due to uncertainties from different models, scenarios, and downscaling methodologies.

The Pan-European Climate Database (PECD4.2), developed in partnership between the European Network of Transmission System Operators for Electricity (ENTSO-E) and the Copernicus Climate Change Service (C3S), marks a change in how energy system planning is conducted.
The PECD4.2 integrates historical reanalysis and six climate models across four SSP's, providing harmonised, openly available datasets tailored for power system studies. 
The physical conversion models for wind and solar energy better reflect technological progression than machine learning methods trained on historical data, improving robustness under changing future conditions.

Despite these advances, challenges remain.
Particularly in hydropower modelling and the lack of public harmonised energy datasets that are required to train these models.
Complex processing chains from raw climate data to actionable insights and the lack of standardized integration of climate information lengthen lead times for energy-sector adoption. 
This leads to diverging approaches and variable consideration of climate risks. 
Closer, more generalised collaboration and communication between climate service providers and energy stakeholders are therefore necessary, as are the development of user-friendly tools for data manipulation and analysis and robust feedback loops. 
\end{abstract}

% Additional keywords
% \vspace{1pc}
\noindent{\it \color{Highlight} Keywords}: Energy Planning; Climate Change; Climate Services; Pan-European Climate Database
\vspace{1pc}

%%%%%%%%%%%%%%%%%%%%%%%%%%%%%%%%%%%%%%%%%%%%%%%%%%
%%%%		~~~~ Introduction ~~~~
%%%%%%%%%%%%%%%%%%%%%%%%%%%%%%%%%%%%%%%%%%%%%%%%%%
\section{Introduction}
The Earth's climate is changing primarily due to anthropogenic greenhouse gas (GHG) emissions and this change is fundamentally transforming the boundary conditions within which energy systems are planned and operated~\autocite{ipcc6wr1}. 
Globally, the energy sector is responsible for approximately two-thirds of GHG emissions~\autocite{IEA2021}. 
A transition toward low-emissions energy sources is therefore critical and advocates for a high level of electrification in energy production. 
A comprehensive decarbonization of electricity production is thus at the forefront of climate change mitigation. 
The Paris Agreement's goals~\autocite{parisAgreement} -- to limit warming to well below 2\textdegree C, preferably 1.5\textdegree C -- demand that future energy pathways are aligned with scenarios which account for a rapidly changing climate, both in average states and in the increasing frequency and severity of extremes.

In this context, the European Green Deal and subsequent legislation (e.g., 'Fit for 55' and REPowerEU) stipulate unprecedented efforts towards a net-zero energy system by 2050, to be achieved chiefly through improved energy efficiency and a major increase in the share of renewables~\autocite{greenDeal,SteppingUp,REPowerEU}. 
This legislation aims for a 55\% reduction of greenhouse gas emissions in 2030 compared to 1990 for the EU27 and to increase the share of renewable energy in the EU's energy mix, with a binding target of 42.5\% for 2030 and the ambition to reach 45\%. 

Electricity systems are therefore expected to play a central role in this transition, notably through the electrification of end uses such as transport, heating, and industry. 
The associated change in electricity demand is expected to be met largely by renewable energy sources, particularly wind, solar photovoltaic, and hydropower. 
This transformation will make energy systems even more dependent on weather, climate variability, and climate change, introducing new vulnerabilities and requiring new tools to assess future risks related to systems' operation~\autocite{Jerez2015,Tobin2016,Tobin2018,Dubus2018,Bloomfield2021,Craig2022}. 

In addition, climate change affects energy infrastructure itself. 
Gradual changes in climate conditions can influence asset performance, efficiency, and lifetime. 
At the same time, extreme events such as heatwaves, floods, droughts, and storms pose growing risks to the reliability of generation, transmission, and distribution assets. 

These trends underscore the necessity to reassess how climate information is integrated into energy system planning and operation. 
This paper addresses the need for a more systematic and harmonised integration of climate change into European energy system studies. 
Building on recent work by European transmission system operators and climate service providers, it reviews the role of climate and energy databases in energy studies (Section~\ref{Sec2}). 
It then presents the main types of energy planning studies and the types of climate data they used (Section~\ref{Sec3}). 
The new Pan-European Climate Database version 4.2 (PECD4.2) is then introduced (Setion~\ref{Sec4}), followed by a discussion on practical applications and remaining challenges toward a wide adoption of a harmonized approach regarding the consideration of climate change in energy prospective studies (Section~\ref{Sec5}).

\section{Climate variability and change as key drivers of power systems}
\label{Sec2}
Energy systems are sensitive to climate across a wide range of temporal and spatial scales. 
Short-term weather variability influences daily and seasonal fluctuations in electricity demand and renewable generation, while long-term climate variability and change affect average conditions, extremes, and the statistical distribution of weather events. 
Energy system assessments require representation and identification of all of these factors, but this is a persistent challenge~\autocite{Craig2022}.  

One of these challenges lies in representing extremes and compound events. 
Periods of low wind and solar generation coinciding with high electricity demand are of particular concern for system adequacy, as they imply a potential energy shortage. 
Climate change may alter the frequency, duration, or intensity of such events, but these changes are difficult to assess using short historical records alone. 
Similarly, compound hazards, such as simultaneous heatwaves and droughts, can affect both electricity demand and hydropower availability~\autocite{Biewald2025}.

Another challenge relates to spatial coherence. 
Large-scale atmospheric circulation patterns can lead to correlated weather conditions across multiple countries, reducing the effectiveness of geographic diversification of renewable resources~\autocite{vanDuinen2025}.
Capturing these correlations requires climate datasets with sufficient spatial coverage and physical consistency.

Finally, uncertainty is inherent to climate information, especially when considering future projections. 
The choice of emission scenarios, climate models, and downscaling methods can substantially influence the projected outcomes. 
Energy system studies must therefore be designed to account for uncertainties, rather than relying on a single deterministic climate future~\autocite{Dubus2022}.

\section{Climate information in energy system studies}
\label{Sec3}
Transmission system operators and policymakers rely on a variety of studies to inform energy planning and investment.

\subsection{Network development and infrastructure resilience}
Network development studies like the Ten-Year Network Development Plan (TYNDP) of ENTSO-E, or similar plans at national level, are core planning instruments that translate long-term energy policy and climate goals into a coordinated infrastructure strategy. 
In Europe they translate high-level policy to targets for 2030, 2040, and 2050. 
They identify power system needs, guide investments in transmission and storage so that the power system remains secure, sustainable and cost-effective across borders. 

Energy infrastructures are long-lived, as much as 20--30 years for solar and wind farms, 50 year or more for nuclear power plants, and up to 80--100 for high voltage overhead lines or hydropower plants. 
Investment decisions in network development should therefore consider also these longer time scales, as it is generally more cost-optimal to design infrastructure that require minimal upgrade during its lifetime. 
Considering future climate change has become critical to limit future adaptation costs to the community.

The 2025 Schéma Décennal de Développement du Réseau (SDDR) by Réseau de Transport d'Electricité (RTE), the French TSO’s national equivalent of the TYNDP, incorporates a strategy to adapt the electricity grid to the foreseeable impacts of climate change in the coming decades. 

The SDDR lays out a gradual, prioritized, and optimized transformation of the electricity transmission system, considering the projected increase in extreme weather events. 
Sensitive sections of the grid, especially in high-risk areas (floods, heatwaves, droughts), will undergo specific reinforcements (infrastructure, equipment, emergency procedures, and monitoring). 
The SDDR aligns with France’s new 2025 National Adaptation Plan for Climate Change, including preparation for a scenario of +2.7\textdegree C warming in France by 2050 and +4\textdegree C by 2100.

In the SDDR assessment several climate databases where used. 
Primarily a 3$\times$200 climate year dataset by Météo-France representing the current climate  and the 2050's under two emission scenarios (RCP4.5 and RCP8.5). 
This was complemented by the DRIAS data to represent the likely climate change by the end of the century~\autocite{soubeyroux2020}. 
Additional datasets from the Copernicus Climate Change Service were also used, especially for anticipating future fire weather risk.

\subsection{Long-term prospective of system operation}
Adequacy assessments aim to evaluate whether available generation and flexibility resources will be sufficient to meet demand at different time horizons, from a season ahead to multi-decadal timescales. 
These are complemented by cost–benefit analyses which focus on optimizing investments into new generation or transmission capacity; operational security analyses aim to ensure a secure delivery of electricity even in case of unplanned outages; market design studies aim to evaluate how reforms could improve power market functioning. 

Traditionally, energy system modelling has utilized historical meteorological datasets, such as the ERA5 reanalysis~\autocite{hersbach2020era5}, to derive input time series for key variables like wind, solar irradiation, and hydropower resources~\autocite{Craig2018,Dubus2018,Bloomfield2021}. 
While these reanalysis products amalgamate observations into consistent, gridded fields suitable for large-scale modelling, they are by construction representative only of past conditions. 
Their validity for supporting long-term adequacy studies diminishes as the climate system departs from historical norms~\autocite{Dubus2022}.

ENTSO-E developed the Pan-European Climate Database (PECD), with versions up to 3.0 based solely on reanalysis products~\autocite{hersbach2020era5}. 
Version 3.1 included only limited ad hoc corrections for climate change, by trend-correcting temperature, and did not directly represent future climate scenarios~\autocite{PECDupdate}.

Adequacy studies are particularly demanding in terms of climate data requirements, as they often rely on long time series of weather-driven demand and generation to assess rare but critical events. 
Climate data used in energy system studies must meet several key requirements (see \textcite{Dubus2022} and references therein). 
First, they should provide long and consistent time series to capture interannual variability and extremes. 
Second, they must preserve the physical relationships between different climate variables, such as temperature, wind, and precipitation. 
Third, spatial and temporal resolution should be sufficient to represent regional differences and relevant processes. 
Finally, transparency and reproducibility are essential, particularly for regulatory and policy-oriented studies.

\subsection{Limitations of current approaches}
In the last few years some TSOs adopted climate projection datasets in their studies. 
These datasets use climate models and GHG emission scenarios to provide potential future climate evolution, most generally up to the end of the century. 
Examples are the SDDR analysis by RTE mentioned above,  their Energy Pathways to 2050 study~\autocite{EnergyPathways2050}, or Elia’s  blueprint for the Belgian electricity system~\autocite{BelgianBlueprint}. 

Despite being a major step compared to the use of historical data, climate scenarios used by early adopters (e.g. RTE and Elia) have some limitations. 
These only used a single climate model, while science recommends considering several models to account for model uncertainty. 
In addition, the fact that the dataset was not publicly available, also limits the transparency and reproducibility. 
Furthermore, the adoption of climate projections remained limited to a few TSOs, which poses a risk of disconnect between national and Pan-European studies.

\section{Pan-European Climate Database version 4.2; a major step forward}
\label{Sec4}
Most climate datasets used until recently presented some limitations, including i) being based mostly on historical data or a single climate model, ii) not being available publicly, iii) not covering the full European domain.

\subsection{The new landscape of climate services in Europe}
Following early development of climate services through European-funded research projects in the late 2000s and early 2010s, the Copernicus Climate Change Service (C3S) was launched in 2015. 
Its goal is to provide authoritative, science-based information on past, present, and future climate change to support policymaking, adaptation, mitigation, and risk management worldwide. 
It aims to turn climate data into actionable knowledge, so societies can understand climate change and respond to it effectively.

The energy sector has been identified as a key sector for C3S applications since the early stages, leading to two proofs of concept (ECEM~\autocite{Troccoli2018,Goodess2019}, CLIM4ENERGY~\autocite{Bartk2019}), followed by a pre-operational service~\autocite{Dubus2023}. 

To respond to the EU Commission and the EU Agency for the Cooperation of Energy Regulators (ACER) ’s request to include climate change in prospective studies, ENTSO-E has partnered with C3S to develop a new, climate-proof version of the PECD~\autocite{Dubus2022}. 

\subsection{PECD4.2, the new reference dataset for ENTSO-E studies}
\begin{figure}[!b]
    \centering
    \includegraphics[width=\linewidth]{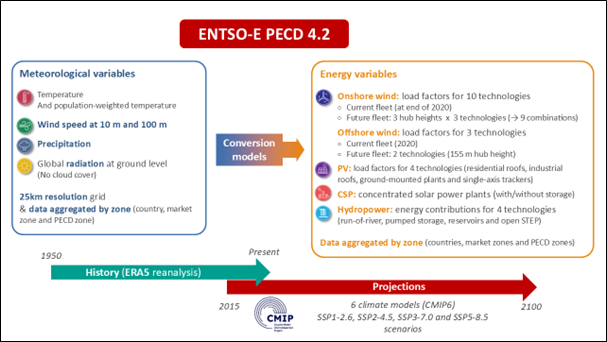}
    \caption{Structure and development of the PECD4.2 in a nutshell}
    \label{fig1}
\end{figure}

Leveraging recent developments in climate services, the PECD has been updated over the last three years to provide a harmonized set of climate-driven energy variables for European transmission system operators (Figure~\ref{fig1})~\autocite{PECD42}. 
It now includes both historical data based on the ERA5 reanalysis~\autocite{hersbach2020era5} and climate projections from 6 CMIP6 global climate models up to 2100, each under 4 Shared Socio-economic Pathways (SSPs) as defined by the IPCC~\autocite{ipcc6wr1}.

The dataset comprises climate variables (air temperature at 2 m, precipitation, solar irradiance, and wind speed) and energy variables related to wind, solar, and hydro generation. 
For each of these, several technologies are considered. 
Onshore and offshore wind are represented by several combinations of hub height and specific power. 
Solar photovoltaic power is split between residential and industrial rooftops, as well as ground-mounted plants, with either fixed orientation or one-axis trackers. 
Hydropower related data include water inflows to reservoirs, pondage, run-of-river and open loop pumping stations. 
Electricity demand is not currently provided but might be included in the next release.

\begin{figure}[!b]
    \centering
    \includegraphics[clip=true, trim=0.2cm 0 0 0, width=\linewidth]{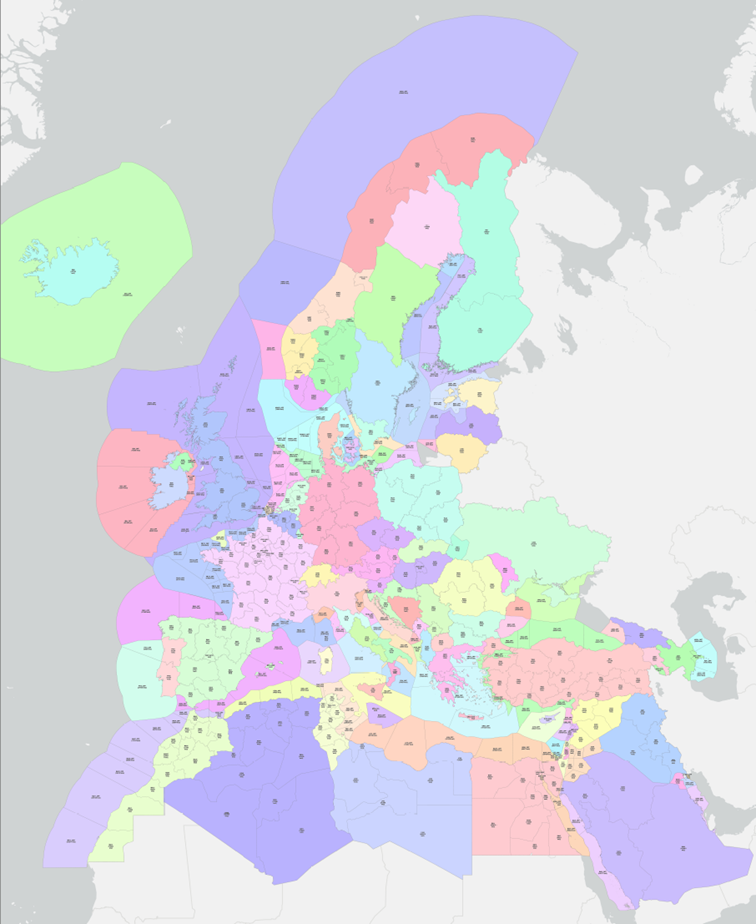}
    \caption{PECD4.2 geographical coverage, sub-national level (PECD zones). Courtesy TenneT and ENTSO-E}
    \label{fig2}
\end{figure}

All climate variables are provided on a 0.25\textdegree×0.25\textdegree grid (approximately 25 km resolution). 
They are also available at different geographic aggregation levels, from sub-national to national scale (Figure~\ref{fig2}). 
Energy variables are provided only at aggregated levels in this version, but future development will also provide gridded energy data.
The temporal resolution is 1 hour for all fields, except precipitation (daily) and hydropower-related data (daily or weekly).

\subsection{Advances within PECD4.2}
The three key advances made within the development of the PECD4.2 are a) the use of physical conversion models for wind and solar to have a robust representation under climate change conditions, b) the incorporation of climate projection-based datasets while maintaining reanalysis datasets for calibration, validation, and bias-adjustment purposes, and c) the explicit consideration of uncertainty through the inclusion of six different climate models and four shared-socio-economic pathways, reflecting uncertainties related to emission scenarios, model structure, and internal climate variability.

In the development of the PECD4.2 particular emphasis was placed on multivariate consistency. 
Even if bias adjustments are made in a univariate approach, the physical and statistical relationships between the climate variables are ensured through the use of physical climate models. 
This is critical for energy system studies that depend on the interaction between demand and multiple renewable generation sources. 
This multivariate coherence is especially important for the analysis of compound events and spatial coherence as outlined in Section~\ref{Sec2}. 

PECD4.2 has been developed in close collaboration between transmission system operators, climate scientists, and climate service providers. 
This co-design process has ensured that scientific robustness is balanced with operational usability. 
Clear documentation, transparent assumptions, and harmonized methodologies, make PECD4.2 a cornerstone for future Pan-European energy system studies under climate change. 

It is planned to use this new database in several studies, notably the ENTSO-E ERAA 2026, the TYNDP 2026, and the ongoing RTE’s update of the Energy pathways to 2050 report. 
Each of these studies has its specific requirements, including differences in the number of climate years that can be processed in their respective power system models (PLEXOS for ENTSO-E, ANTARES simulator for RTE). 
As a result, climate years subsets selection methodologies have also been proposed~\autocite{vanDuinen2026} and will be further developed in the coming months. 
These aim to identify representative years for a given period, while preserving as much of the variability in the full data ensemble as possible and allowing more explicit alignment with the different study objectives.

\clearpage
\subsection{Limitations}
PECD4.2 is a major step forward compared to previous databases. 
However, some further developments are still required. 
These include improving the energy conversion models, particularly for hydropower, which currently rely on basic machine learning models (random forests) using only temperature and precipitation as inputs. 
A physical hydrological model could address this limitation and is currently being considered for inclusion in the next version of the PECD.

The key limitation of the PECD4.2 is the focus of the generality of the conversion models over accuracy for each country. 
Combined with the use of the ENTSO-E Transparency Platform data for calibration and validation this has led to some quality issues mainly related to data availability. 

\section{Discussion and steps forward}
\label{Sec5}
Despite significant progress several limitations remain in the use of climate databases for prospective energy studies. 
For the PECD4.2 further work is planned through a new C3S contract to start in 2026, that will try to address its current limitations and bring additional improvements.

In general the development and evaluation of climate-energy conversion models are often hampered by the limited availability and, at times, uneven quality of the necessary energy data (e.g., insufficient temporal and spatial scales, restricted access, verification issues). 
Further analysis of the potential improvements enabled by higher-quality energy data could help drive collective data-gathering efforts within the TSO community.

The data processing chains from climate-energy data generation to practical use in energy prospective studies remain complex, which lengthens the time between the production of climate simulations (CMIP, CORDEX), their processing by climate services (C3S), and their adoption and use in the energy sector. 
As part of the PECD4.2 development, a set of Python notebooks~\autocite{PECDtraining} is provided, offering a good starting point for non-specialist users to access, retrieve, and process the data. 
It will be further developed in the future.

Managing uncertainties is a real challenge for practical applications: selecting climate years for energy models requires representative sampling methods that do not underestimate variability or extreme events. 
Furthermore, from a sectoral perspective, climate change is currently considered unevenly, depending on the actors involved. 
In particular, standards relating to materials and methodologies are disparate. 
They should be reviewed, not only to ensure that climate change is better considered, but also to promote greater consistency in approaches between actors. 

To overcome these obstacles, it is essential to strengthen collaboration between climate service developers and energy users, and to provide tools that facilitate data manipulation, analysis, and visualization. 
The development of an effective feedback loop and the widespread availability of detailed energy datasets are key factors in increasing the relevance and robustness of the models and methods used in prospective studies.

Despite some limitations, PECD4.2 marks a step change by offering a science-based, user-driven reference database to all actors in the energy sector.

%%%%%%%%%%%%%%%%%%%%%%%%%%%%%%%%%%%%%%%%%%%%%%%%%%
%%%%		~~~~ Backmatter ~~~~
%%%%%%%%%%%%%%%%%%%%%%%%%%%%%%%%%%%%%%%%%%%%%%%%%%
\newpage
% Contribution statement
\section*{CRediT Author Statement}
Conceptualization of PECD4.2: LD \& LS with support from ENTSO-E Expert Team Climate, Development and production of PECD4.2: AT with support from ICS and C3S, Writing - Original Draft: LD, Writing - Review \& Editing: All authors. 

% Acknowledgement section
\section*{Acknowledgments}
The content of this paper and the views expressed in it are solely the author’s responsibility, and do not necessarily reflect the views of TenneT, RTE, or, ENTSO-E or its members.

% Statement on open & FAIR research
\section*{Open research} 
The PECD4.2 is available under the as the \emph{Copernicus Climate Change Service (2024): Climate and energy related variables from the Pan-European Climate Database derived from reanalysis and climate projections}.
For more details, download options and the full documentation see \textcite{PECD42} at \url{https://doi.org/10.24381/cds.f323c5ec}

% Bibliography / References
\printbibliography

\end{document}